\documentstyle[preprint,aps]{revtex}
\tighten

\begin{document}
\draft
\preprint{LPTENS 94/03}
\title{ Effect of a magnetic field on Mott-Hubbard systems}
\author{Laurent Laloux*, Antoine Georges* and Werner Krauth**}
\address{
* CNRS-Laboratoire de Physique Th\'{e}orique de l'Ecole Normale
Sup\'{e}rieure$^1$
\\
e-mail: laloux@physique.ens.fr, georges@physique.ens.fr
\\
*** CNRS-Laboratoire de Physique Statistique de l'Ecole Normale
Sup\'{e}rieure$^2$
\\
e-mail: krauth@physique.ens.fr
\\
24, rue Lhomond; 75231 Paris Cedex 05; France}

\date{\today}
\maketitle
\begin{abstract}
The effect of a magnetic field on Mott-Hubbard systems is investigated
by studying the half-filled Hubbard model in the limit of infinite
dimensions. A first-order metamagnetic transition between the
strongly correlated metal and the Mott insulator is found for a
critical value of the applied field. The field and temperature
dependence of the magnetization, one-particle properties and
susceptibility are studied and compared to the Gutzwiller
approximation. The experimental relevance
for transition-metal oxides and liquid $^{3}He$
is discussed.

\vspace{3cm}
\begin{description}
 \item[$^1$]  Unit\'{e} propre du CNRS (UP 701) associ\'{e}e \`{a} l'ENS  et
\`{a} l'Universit\'{e} Paris-Sud
 \item[$^2$]  Laboratoire associ\'{e} au CNRS (URA 1306) et aux
Universit\'{e}s Paris VI et Paris VII
\end{description}
\end{abstract}

\narrowtext
\section{Introduction}

Systems which are close to a Mott-Hubbard transition between a paramagnetic
metal and a paramagnetic insulator display local magnetic moments interacting
through a residual antiferromagnetic exchange. Hence, the response of
such systems to a
magnetic field is an interesting physical problem in which a competition
takes place between the exchange and the alignment of the local moments with
the external field. On the metallic side of the transition, the problem is
even more involved since an additional energy scale exists (the
local Kondo temperature, or effective Fermi energy) associated with the
quenching of local spin fluctuations at low temperature.

Up to now, the only available quantitative description of this problem has
been the Gutzwiller approximation \cite{GA,BR,DV,PN}.
The susceptibility of the strongly correlated metal is predicted
to increase with the field in this approach,
and a first-order localization
transition (metamagnetic transition) is found for a critical value
of the applied field
\cite{DV,PN}. The Gutzwiller approximation has limitations however.
The main one is that it neglects the residual
exchange altogether, so that it cannot account for the physics of the above
competition. The Mott insulator is caricatured
as a collection of independent local moments which has infinite
susceptibility at zero temperature, and the susceptibility
also diverges as the metal/insulator
transition is reached from the metallic side.
Finally, this approximation is restricted to
zero temperature (although some finite-temperature extensions
have been attempted \cite{PN,SG,KR}).

Recently, a new approach to the Mott transition has been proposed
\cite{AW1,RZK} and
extensively studied \cite{AW2,ZRK,RKZ,RMK,MSKR},
based on the Hubbard model in the controlled limit
of infinite spatial dimensionality ($d\rightarrow\infty$)
\cite{MV,GK,MJ}. The exchange is (at least partially) taken into account
in this limit,
and finite temperature effects can be addressed in a consistent manner.
The aim of the
present paper is to study the effect of a magnetic field on the
Mott transition (as
well as on the correlated metal or deep into the insulator) within
this approach. Some earlier attempts have appeared in the literature
\cite{AW2,RKZ,SH}, but
have not been able to solve the problem in the
low-temperature or low magnetic field regime because of limitations in
the numerical method employed. We find that at very low
temperature a magnetic field does drive the strongly correlated
metal closer to localization, and that a first-order metamagnetic transition
to the Mott insulator takes place for a critical value of the applied field.
This is our main new result, which is in qualitative agreement with the
predictions of the Gutzwiller approximation (even though the magnetic
exchange significantly modifies  the quantitative results
of this approximation).
We establish that, near this transition and at finite
temperature, the magnetic susceptibility is an {\it increasing} function
of the magnetic field.
We also provide further evidence that the
zero-field Mott transition is first-order at finite temperature,
and show that the magnetic properties of both the weakly correlated metal
and the Mott insulating phase can be understood quantitatively in a simple
manner.

There are several experimental motivations to our work. It has been
demonstrated in previous studies \cite{AW2,RKZ} that the $d=\infty$ approach
to the Mott transition agrees qualitatively (and, for some properties,
quantitatively) with many observed features of the paramagnetic
metal/paramagnetic insulator transition of transition metal oxides.
There have been some investigations of magnetic properties close to this
transition (e.g for $(V_{1-x}Cr_x)_2O_3$ in ref.\cite{MR}), with which
our results are in satisfactory qualitative agreement.
The recent experiments on the field dependence of the magnetization of
liquid $^3He$ \cite{WWP} provide another important motivation.
There are two competing descriptions of this
strongly-correlated Fermi liquid.
The Stoner and paramagnon approach \cite{PMAG} views
the system as close to a ferromagnetic transition while the
`almost localized' approach \cite{AB,DV} views it as a
strongly-correlated liquid close
to Mott localization, and relies on a Hubbard model lattice description.
Quantitative predictions in the latter approach have up to now relied on
the use of the Gutzwiller approximation \cite{DV}.
These two approaches have
led to very different predictions
for the response to an external field:
the Stoner approach predicts a smooth magnetization
($\partial \chi/\partial h <0$), while
the Gutzwiller approximation exhibits a first-order metamagnetic transition,
with $\partial \chi/\partial h >0$ close to the transition.
One of the main purpose
of the above experiments has been to discriminate between them.
Our results on a well-controlled limit of a specific model
provide a test of both approximation schemes and confirm that
a description of liquid $^3He$ by a lattice Hubbard model
rigidly maintained at half-filling is inappropriate.

This paper is organized as follows. In section II we define the model,
explain the numerical methods and briefly summarize previously
established results on
the zero-field Mott transition in the $d=\infty$ limit. In section III, we
give an overview of our numerical results for the field dependence
of the magnetization and of the phase diagram as a function of field and
temperature. In section IV, some aspects of the Gutzwiller approximation
for the model under study are summarized. Section V is devoted
to a detailed description
and discussion of our results, and section VI to some comparisons with
experiments and concluding remarks.

\section{Model, Methods, and Zero-field Mott Transition}

\subsection{The Hubbard Model in infinite Dimensions}

\noindent
We consider in this paper the Hubbard model in a uniform magnetic field
at half-filling \mbox{($\mu=U/2$)}:

\begin{equation}
H = - \sum_{<ij>,\sigma} (t_{ij} c^{+}_{i\sigma} c_{j\sigma} + h.c) + U
\sum_{i} n_{i\uparrow} n_{i\downarrow}
- \sum_{i \sigma} (\mu +h\sigma)  n_{i\sigma}
\end{equation}
with nearest-neighbor hopping on a lattice of connectivity $z$,
in the limit $z\rightarrow\infty$.
The hopping must be scaled \cite{MV} as $t_{ij}=t/\sqrt{2 z}$ to keep
the problem non-trivial in this limit. For simplicity, we  consider in
the following a Bethe lattice, for which
the free ($U=0$) density of states (d.o.s.)
\mbox{$D(\epsilon)=\sum_{{\bf k}} \delta(\epsilon-\epsilon_{\bf k})$}
takes a semi-circular form in the limit $z\rightarrow\infty$:
$D(\epsilon)=\frac{1}{\pi t} \sqrt{2-(\epsilon/t)^2}$.
Unless explicitly stated, we set $t=1$ in the following.

Following ref.\cite{GK} ({\it cf}  also \cite{BM,VJ}),
all the local properties
of the model can be obtained {\it via} a single-site impurity problem
supplemented by a self-consistency condition.
The effective action of this impurity problem, in the presence of a
magnetic field, reads:
\begin{equation}
\cal{S} = - \int^{\beta}_{0} d\tau \int^{\beta}_{0} d\tau'
\sum_{\sigma} c^{+}_{\sigma}(\tau)[G_{0 \sigma}^{-1}(\tau - \tau')]
c_{\sigma}(\tau')
+ U \int^{\beta}_{0} d\tau n_{\uparrow}(\tau) n_{\downarrow}(\tau)
\label{S}
\end{equation}
where, in the {\em paramagnetic phase},
the `bare' Green's function $G_{0 \sigma}$
is related to the interacting Green's function
$G_{\sigma}(\tau-\tau')\equiv -<Tc(\tau)c^{+}(\tau')>_{\cal{S}}$
through a self-consistency equation. For the Bethe lattice, it reads:
\begin{equation}
G^{-1}_{0 \sigma}(i\omega_{n})  = i\omega_{n} + \mu + h\sigma -
\frac{t^{2}}{2} G_{\sigma}(i\omega_{n})
\label{sc}
\end{equation}
The self-energy of the lattice model reads:
$\Sigma_{\sigma}(i\omega_{n})=G_{0\sigma}^{-1}(i\omega_n)-
G_{\sigma}^{-1}(i\omega_n)$ and is independent of momentum \cite{MH} in
the $z\rightarrow\infty$ limit.
Because of the magnetic field, these equations depend on  spin, but at
half-filling (\mbox{$\mu = U/2$}), the symmetry property
$F_{\sigma}(- i\omega_{n}) = - F_{- \sigma}(i\omega_{n})$
holds, where
$F$ is any of the functions:
$G_{\sigma}, G_{0\sigma} - \mu, \Sigma_{\sigma} - \mu$.

It is useful \cite{GK} to view the effective action $\cal{S}$ as that of an
Anderson model
with a spin-dependent hybridization function $\Delta_{\sigma}(\omega)$,
where the conduction electron `bath' has been integrated out:
\begin{equation}
G^{-1}_{0 \sigma}(i\omega_{n}) = i\omega_{n} + \mu + h\sigma -
\int^{+ \infty}_{- \infty} d\omega\,\,
\frac{\Delta_{\sigma}(\omega)}{i\omega_{n} - \omega}
\end{equation}
For the Bethe lattice, the self-consistency condition (\ref{sc})
specifies the hybridization function in terms of the
interacting local spectral density
$\rho_{\sigma}(\omega)\equiv -\frac{1}{\pi} Im G_{\sigma}(\omega+i0^+)$
through:
$\Delta_{\sigma}(\omega) = \frac{t^{2}}{2} \rho_{\sigma}(\omega)$
Although highly simplified, this coupled problem remains unsolvable
analytically:
numerical methods are necessary to obtain a full solution.

In this work, we follow the {\it paramagnetic} solutions of these
coupled equations, even though the model actually has a transition to
an antiferromagnetic phase below some N\'{e}el temperature for arbitrary
$U$ (the Bethe lattice is a bipartite lattice). This is possible since
$z\rightarrow\infty$ is a mean-field limit in which the
continuation of a solution of the mean-field equations in an unstable
region has a well-defined mathematical meaning. Alternatively, one may
wish to consider a model which does not display antiferromagnetic
ordering and for which the solutions studied in this paper describe the
actual ground-state.
As discussed in ref.\cite{AS,AW2,RKZ}, an example of such a model is
a Hubbard model on a {\it fully connected} cluster of $N$ sites,
with randomness on the hopping parameters $t_{ij}$:

\begin{equation}
H = - \sum_{\sigma;i,j=1}^{N} t_{ij} c^{+}_{i\sigma} c_{j\sigma} + U
\sum_{i} n_{i\uparrow} n_{i\downarrow}\,\,\,,\,\,\,
with: \overline{t_{ij}^2}={{t^2}\over{2N}}
\label{rm}
\end{equation}
It can be shown that the
single-particle Green's function of this random model coincides with the
same quantity for the {\it paramagnetic phase} of the Hubbard
model on the $z=\infty$ Bethe lattice,
with no randomness and hopping parameter $t_{ij}=t/\sqrt{2z}$.
It is clear however that the phase
diagram of the two models differ: the disordered one has a highly
degenerate singlet ground state for large $U$ at half-filling, and
{\it no antiferromagnetic phase}.

\subsection{Numerical Method: exact Diagonalization for a fixed
Magnetization}

For a given value of the interaction strength and temperature,
and a given magnetic field $h$, the above set of coupled equations
can be solved following the usual two-step procedure:

(i) calculation of $G_{\uparrow},G_{\downarrow}$ for a given pair of
$G_{0\uparrow},G_{0\downarrow}$

(ii) calculation of $G^{new}_{0\uparrow},G^{new}_{0\downarrow}$
using the self-consistency relation.

\noindent
Step (i) involves the solution of the Anderson model
with a given
conduction `bath'.
This can be achieved by two possible algorithms: the Quantum Monte Carlo
Hirsch-Fye algorithm, first used in this context in ref.\cite{MJ,AW1,RZK}
and the exact diagonalization method recently introduced
\cite{MW}.
As shown in previous work, the exact diagonalization method is far more
efficient for the problem at hand than Quantum Monte Carlo, against
which it has been carefully checked,
and we shall use it in the present work.
Step (ii) updates the conduction bath self-consistently.
An iteration of this procedure indeed converges for not too small
magnetic fields and not too close to the metal/insulator transition.
In order to investigate these ranges of parameters, a modification of
this procedure must be used \cite{foot1}, which seeks convergence
for a {\it fixed} value
of the {\it magnetization}, as  described below. This is the main
technical point which allows us to obtain the new results described
here.

The continuous conduction electron bath is parametrized by a
finite number of parameters
(\mbox{$ \{\epsilon_{k \sigma} , V_{k \sigma}\} , k=(2,\cdots ,n_{s}) ,
\sigma = \uparrow,\downarrow$}).
The Hamiltonian of the corresponding  Anderson model reads:
\begin{equation}
H_{And} = \sum_{\sigma} \epsilon_{d \sigma} d^{+}_{\sigma} d_{\sigma} +
\sum^{n_{s}}_{\sigma , k=2}
\epsilon_{k \sigma} a^{+}_{k \sigma} a_{k \sigma}
+ U n_{d \uparrow} n_{d \downarrow}
+ \sum^{n_{s}}_{\sigma , k=2}
( V_{k \sigma} a^{+}_{k \sigma} d_{\sigma} + h.c.)
\end{equation}
The Green's function $G_{0\sigma}$ of the impurity site $d_{\sigma}$
is then represented by:
\begin{equation}
G^{And}_{0\sigma} = [i\omega_{n} - \epsilon_{d \sigma}
- \sum^{n_{s}}_{k=2} \frac{V^{2}_{k \sigma}}{i\omega_{n}
- \epsilon_{k \sigma}}]^{-1}
\end{equation}
where $i\omega_{n}=(2n+1)\pi/\beta$,
and with the  symmetries at half-filling:
\begin{equation}
\left\{ \begin{array}{ccl}
\epsilon_{d \sigma} & = & - ( U/2 + h \sigma ) \\
 \epsilon_{k \downarrow} & = & - \epsilon_{k \uparrow} \\
V^{2}_{k \downarrow} & = & V^{2}_{k \uparrow}
\end{array}
\right.
\end{equation}
An exact diagonalization of $H_{And}$ is performed to obtain the
Green's
function $G_{\sigma}$, and a function $G^{new}_{0\sigma}$ is deduced
from the self-consistency condition.
Then, the new Green's function $G^{And \ new}_{0\sigma}$ is calculated.
This is done by minimizing with respect to
\mbox{$\{\epsilon^{new}_{k \sigma} , V^{new}_{k \sigma}\}$} the
following cost function, {\em on the imaginary axis},

\begin{equation}
\chi^{2} = \frac{1}{n_{max}+1} \sum^{n_{max}}_{n=0}
|G^{-1 new}_{0\sigma}(i\omega_{n})
- G^{-1 And \ new}_{0\sigma}(i\omega_{n})|
\end{equation}
In practice, a conjugate gradient method is used to perform the
minimization. On a workstation, complete diagonalization of the
Hamiltonian is possible for $n_{s} \leq 6$ at all temperatures,
and the Lancz\`{o}s algorithm allows us to calculate directly
zero-temperature properties up to $n_{s} \sim 10$.
At zero temperature, $\beta$ is used as a fictitious temperature serving
as a cutoff at small frequencies.
The difference between $G_{0\sigma}$ and $G^{And}_{0\sigma}$
provides a test of the accuracy of the method (which is exact in the
limit $n_s\rightarrow\infty$). In practice, a good accuracy is obtained
already for $n_{s} = 4$ for a temperature as low as $\beta \simeq 100$.

We now describe in more detail the precise algorithm used
at a {\it fixed value of the magnetization}:

(i) for a given set $\{\epsilon_{k \sigma} , V_{k \sigma}\}$ , we find the
magnetic field $h = -\mu - \epsilon_{d \uparrow}$ which gives the desired
magnetization $m$.
Then we calculate the Green's functions $G_{\sigma}$ and $G^{new}_{0\sigma}$
for this field $h$.

(ii) the new set $\{\epsilon^{new}_{k \sigma} , V^{new}_{k \sigma}\}$ is
obtained from $G^{new}_{0\sigma}$ with the $\chi^{2}$ fit.

\noindent
The convergence is obtained for a fixed value of the interaction $U$ and the
magnetization $m$, when both \mbox{$\{\epsilon_{k \sigma} , V_{k \sigma}\}$}
and $h$ are stabilized.

\noindent
Using the above method, we are interested in computing the following physical
quantities:

{\em Uniform magnetization}, defined as:
\begin{equation}
m = < n_{\uparrow} - n_{\downarrow} > = G_{\uparrow}(\tau = 0^{-}) -
G_{\downarrow}(\tau = 0^{-})
\end{equation}

{\em Uniform Susceptibility} $\chi$, calculated by
numerical differentiation:
\begin{equation}
\chi = \frac{m(h+\Delta h)-m(h)}{\Delta h}\,\,,\,\,\Delta h\rightarrow 0
\label{chi}
\end{equation}
Note that the susceptibility can also be calculated from an evaluation
of a 2-particle Green's function for the impurity model \cite{ZL,MJ}.

{\em Local Susceptibility} The local susceptibility $\chi_{loc}$ is the
response of the system to a local magnetic field applied
{\it on a single site} (taken as the impurity site):
\begin{equation}
\chi_{loc} = \int^{\beta}_{0} d\tau <T [(n_{\uparrow}(0) -
n_{\downarrow}(0)) - m]
[(n_{\uparrow}(\tau) - n_{\downarrow}(\tau)) - m]>
\end{equation}
\begin{equation}
=\sum_{\alpha,\gamma}
{{|<\alpha|S_z-m|\gamma>|^2}\over{E_{\gamma}-E_{\alpha}}}
(e^{-\beta E_{\alpha}}-e^{-\beta E_{\gamma}})
\end{equation}

{\em The local density of states}
$\rho_{\sigma}(\omega)=-{{1}\over{\pi}}Im G_{\sigma}(\omega+i0^+)$
can be computed
from the spectrum and matrix elements.
For finite $n_s$, it is a sum of delta-functions which nevertheless give
a satisfactory account of the main spectral features.

\subsection{The Mott Transition in Zero-field}

We briefly summarize in this section the understanding of the
Mott transition for zero external field that has been reached in previous
work \cite{AW1,RZK,AW2,ZRK,RKZ,RMK,MSKR} and which is further
extended by the present study.
The coupled equations  (\ref{S},\ref{sc}) for the zero-field Green's
function have two types of solutions at zero temperature. Metallic
solutions have the characteristic low-frequency behavior of a Fermi liquid
\cite{foot2}:
\mbox{$Re  \Sigma(\omega+i0^{+})=U/2+(1-1/Z)\omega+... $},
\mbox{$Im \Sigma(\omega+i0^{+})=-\Gamma\omega^2+... $} .
$Z$ is the quasiparticle residue, which is related here
to the effective mass by
$m^*/m=1/Z$ \cite{MH}.
A plot of $Z$ as a function of $U$ is given in the inset of
Fig. \ref{chilocvsu}.
In contrast, insulating solutions have
\mbox{$Re  \Sigma(\omega+i0^{+})=C/\omega+...$} at low frequencies, while
$Im\Sigma$ and $\rho(\omega)$  vanish (except for a $\delta(\omega)$ piece
in $Im\Sigma$) inside a finite
frequency range $[-\Delta_g/2,\Delta_g/2]$.
$\Delta_g$ is the Mott gap to
charge excitations, and the effective mass is infinite ($Z=0$).

In the metal, a local moment exists at high temperature (with the local
susceptibility $\chi_{loc}$ following Curie's law) but is quenched
at low temperature, so that $\chi_{loc}$ is finite at
$T=0$. This quenching is associated with the Kondo effect of the
associated impurity
model, since for metallic solutions the conduction electron `bath'
has a finite density of states at the Fermi level ($\rho(0)\neq 0$).
The spin-fluctuation
energy scale corresponding to this quenching is the local Kondo temperature
$T_K$,
of the order of $Z t$. The conduction electron `bath' d.o.s. of insulating
solutions has zero spectral weight at low frequency, so that the Kondo
effect does not take place and an unquenched local moment exists down
to zero temperature. As
a result $\chi_{loc}$ follows Curie's law and diverges at $T=0$.

A plot of the local susceptibility
$\chi_{loc}$ {\it vs.} interaction strength is displayed in
Fig. \ref{chilocvsu} for a finite but low temperature $T=0.01$.
It is immediately apparent on this plot that a regime of
{\it coexisting solutions} exists
for $U_{c1}(T)<U<U_{c2}(T)$, and thus that the Mott transition
is a {\it first-order transition} at finite temperature (as in a liquid-gas
system), as
previously suggested in Ref.\cite{AW2,RKZ}.
(The study of the coexistence interval with $n_{s}$=3, 4, 5, 6 gives the
{\it stabilized}
values: $U_{c1} \simeq 3.3$ and $U_{c2} \simeq 3.8$ at $T=0.01$ ).
At finite temperature
entropic effects strongly
favor the `insulator' (which has a ground-state entropy $Nln2$ in this
model) and the first-order transition line must therefore be
close to $U_{c1}(T)$.

At zero temperature, a locally stable metallic solution exists for
$U<U_{c2}(0)$, while an insulating solution is found for $U>U_{c1}(0)$.
Whether $U_{c2}(0)$ is actually different from $U_{c1}(0)$
has not yet been entirely clarified. Evidence that
$U_{c1}(0)\simeq U_{c2}(0)$ has been provided by the Lancz\`{o}s results of
ref.\cite{MW}. Another recent study \cite{RMK}
concludes also that the $T=h=0$ transition is second-order and takes place
at $U_{c2}(0)$ .
These critical points are {\it a priori} associated with quite
different physical phenomena however.
$U_{c2}(0)$ is a second-order critical point \cite{ZRK,MSKR}
at which $m^*/m$, $\chi_{loc}$
and the inverse compressibility diverge,
much as in the Brinkman-Rice
scenario \cite{BR}.
This reflects the continuous disappearance of coherent low-energy
excitations (quasiparticles) in this almost localized Fermi liquid.
On the other hand, the disappearance of the insulating solution as $U$
is lowered below
$U_{c1}(0)$ occurs when the gapped conduction electron `bath' can no longer
sustain an unquenched local moment. Whether this happens abruptly or
with a gap closing continuously at $U_{c1}(0)$
has not yet been settled in this model.

The response of these phases to a small magnetic field has been discussed
in previous work ({\it cf} \cite{AW2} and  \cite{RKZ}).
The inset of Fig. \ref{chilocvsu}
displays numerical results for the uniform magnetic susceptibility
$\chi$ as a function of $U$, obtained
at $\beta=100$.
$\chi$ is finite down to $T=0$ in
both phases and {\it does not diverge} as $U_{c2}(0)$ is approached within
the metallic solution, in contrast to $\chi_{loc}$. The reason is that the
energy scale for residual antiferromagnetic exchange entering {\it local}
response functions (such as $\chi_{loc}=\sum_{q}\chi(q)$) is the exchange
between two {\it fixed} sites $J_{ij}=t_{ij}^2/U=O(1/d)$, while the
scale entering uniform response functions (such as $\chi=\chi(q=0)$) is
the sum of $J_{ij}$ over all neighbors: $J=t^2/2U$. Hence the former scale
disappears from the physics as $d\rightarrow\infty$, while the latter
remains $O(1)$: the exchange is not treated
on equal footings in uniform and local quantities in this limit. This
explains also why the $d=\infty$ Mott insulating phase has a finite
ground-state
entropy $Nln2$ but a finite uniform magnetic susceptibility
($\chi=1/J$ for large $U$).

\section{Overview of the Results for finite Field}

Figure \ref{mag(uvar)} displays our numerical results
for the field dependence of the magnetization, for various values of
$U$ and
a fixed temperature $T=0.01$. Three different behaviors are clearly
seen. For the smaller values of $U$, a unique metallic solution is
found in which the magnetization smoothly saturates as the field is
increased. For the larger values of $U$, a unique insulating solution
is found in which the magnetization quickly saturates (corresponding to
a large $\chi(h=0)$), and a rapid crossover from the Mott insulator to
the fully polarized band insulator is found. The most interesting
behavior arises for intermediate values of $U$, an example of which is
$U=3$ on Fig. \ref{mag(uvar)}. There, two different solutions
are found, one at lower field is 'metallic' (for a more precise
characterization, see section V), and disappears
at some critical field $h_{c2}(U,T)$ and the other is 'insulating' and
only exists at fields larger than $h_{c1}(U,T)$. We find for $U=3$ and
$T=0.01$ that a coexistence region exists, {\it i.e.} that
$h_{c1}\simeq 0.13 <h_{c2}\simeq 0.17$. Note that the value $U=3$ is
{\it below} the critical value $U_{c1}(h=0,T=0.01)$ above which an
insulating solution is found in zero field (this is why $h_{c1}>0$ in this
case). Hence a magnetic field
drives the strongly correlated metal to a first-order metal/insulator
transition at some critical field at which the magnetization jumps
discontinuously (metamagnetic transition).
It plays in this respect a role similar to temperature,
which also drives the Mott transition first-order.
We have also calculated magnetization curves for $U=3.5$, which lies
inside the zero-field coexistence region at $T=0.01$,
so that both a metallic and an insulating solution exist for small fields.

Based on these findings, we can draw an estimate of the
phase diagram in the $(U,h)$ plane at
$T=0.01$, as shown in the inset of Fig. \ref{mag(uvar)}.

One can also follow the temperature
dependence of these solutions, as shown in Fig. \ref{mag(betavar)} for
$U=3$. It is only at the lowest temperatures studied that the coexistence
sets in. For temperatures $T$ above roughly $1/50$ in this case, a single
solution is recovered. For this reason, a detailed investigation of this
phenomenon is beyond the reach of present  Monte Carlo methods and has
escaped previous studies \cite{AW2,RKZ,SH}.

\section{Magnetic Properties in the Gutzwiller Approximation}

For further comparison with our results, we review in this section
some aspects of the response to a magnetic field of the Hubbard model
within the Gutzwiller approximation.

The Gutzwiller method \cite{GA,BR,DV,PN} is a zero-temperature
variational approach which relies on the ground-state wave function:
$g^{D} |FS>$, where $|FS>$ denotes a free Fermi-sea, $D$ is the
double occupancy operator
$D = \sum_{i} n_{i \uparrow} n_{i \downarrow}$, and $0 \leq g \leq 1$
is a variational parameter.
In the limit of infinite dimensions, average values computed with
this wave function coincide with the results of the `Gutzwiller
approximation', so that no distinction needs to be made between the
two approaches \cite{MV}.

At half-filling and for finite $h$, the variational ground-state
energy reads (using \mbox{$d\equiv<n_{\uparrow} n_{\downarrow}>$}
as a variational parameter instead of $g$):

\begin{equation}
E_{g}=4d{{1-2d+\sqrt{(1-2d)^2-m^2}}\over{1-m^2}} \epsilon_0(m) +U d
\end{equation}
where:

\begin{equation}
\epsilon_0(m)=\int_{-\infty}^{\mu_0(n_{\uparrow})}d\epsilon\  \epsilon
D(\epsilon)
+\int_{-\infty}^{\mu_0(n_{\downarrow})}d\epsilon\  \epsilon
D(\epsilon)
\end{equation}
The corresponding applied field is obtained from:
$h =  \ \frac{\partial E_{g}}{\partial m}$.

For zero external field, the solution of this
variational problem
predicts \cite{BR} a second-order Mott transition at
the Brinkman-Rice point
\mbox{$U_{BR} = \frac{32 \sqrt{2}}{3 \pi}  \simeq 4,8 $}
between a metallic phase for $U<U_{BR}$ and an insulating phase
for $U>U_{BR}$.
This transition is characterized by the divergence of the effective mass
($Z=m/m^{*} = 1-(U/U_{BR})^{2}$), the susceptibility
($\chi /\chi_{0} \propto m^{*}/m$) and the inverse of the compressibility.
The gap of the insulating solution opens up continuously at $U_{BR}$:
$\Delta_g\propto (U-U_{BR})^{1/2} $.
Because the exchange is neglected in this description, the insulator is
just a collection of independent local moments: the ground-state entropy
is $Nln2$, and {\it both} the
local and uniform magnetic susceptibility are infinite at zero temperature
(in contrast to the $d=\infty$ description which has $\chi_{loc}=\infty$
but $\chi$ finite). Finite temperature extensions of the
Gutzwiller approximation, following
{\it e.g.} the four-slave-boson scheme \cite{KR},
predict a first-order transition for $T>0$ \cite{PN}.

The effect of a uniform  magnetic field $h$ within the Gutzwiller
approximation has been addressed in detail by
Vollhardt \cite{DV} and by Nozi\`{e}res \cite{PN}.
For small $U$, the magnetization of the metal smoothly saturates with
$\partial \chi / \partial h < 0$.
But, for \mbox{$U \geq .44 U_{BR}$},
the metal becomes metamagnetic with $\partial \chi / \partial h > 0$.
Furthermore, close enough to $U_{BR}$, $Z$ {\it decreases} upon
increasing $h$: the field drives the system closer to localization.
A general thermodynamical identity (`Maxwell relation') relates the
specific-heat enhancement to the low-temperature behavior of the
susceptibility:

\begin{equation}
\frac{\partial^{2} \chi}{\partial T^{2}} =
\frac{\partial^{2} C_{v}/T}{\partial h^{2}} \equiv
 - \frac{2 \pi^{2}}{3} D(0) \
\frac{1}{Z(h=0)^{2}} \ \frac{\partial^{2} Z}{\partial h^{2}}
\label{Maxwell}
\end{equation}
where all derivatives are taken at $h=T=0$.
This implies \cite{PN} that any
consistent
finite-temperature extension of the Gutzwiller approximation would yield
$\partial \chi / \partial T >0 $ close to $U_{BR}$.

Furthermore, for sufficiently large $U$, the variational equations have
three solutions in a certain range of values of $h$,
one of which is unstable thermodynamically.
A mixture of the two stable solutions may lower the energy, and
a Maxwell construction must be made to find the field at which
a first-order
transition takes place \cite{PN}.
The resulting magnetization curves $m(h)$ are depicted in
Fig. \ref{mGutzvsu} for various values of $U$.
Hence, a line of first-order phase transition is found in the plane $(U,h)$,
as shown in  the inset of Fig. \ref{mGutzvsu}, ending at a second-order
point $(U_{c},h_c)$. Correspondingly, a
coexistence region can be drawn in the $(U,m)$ plane:
phase separation takes place for $m_{c1}(U)<m<m_{c2}(U)$ when $U>U_c$.

\section{Results and Discussion}
\subsection{The weakly correlated Metal}

We consider first the metallic phase with moderate correlation effects
($U$ much smaller than $U_{c1}$). There,
the effect of a magnetic field is to drive the system from the unpolarized
to the fully magnetized Fermi liquid, which at half-filling is
actually a band insulator. This process is a smooth crossover.
Because of the Pauli
principle, the polarization of the spins makes the interaction term
\mbox{$n_{\uparrow}n_{\downarrow}$} less and less effective as the field is
increased. Accordingly, the interaction-enhanced effective mass decreases
smoothly towards $m^*=m$ as $h$ increases,
as depicted in the inset of Fig. \ref{mvsh(u=1)}, where $Z=m/m^*$ is
plotted $vs. h$ for $U=1$.
The crossover scale is the effective Fermi energy, or
spin-fluctuation Kondo scale $T_K$. The Fermi liquid (metallic) character
of the solution all the way to the fully polarized state is demonstrated by
checking that the solution always satisfies Luttinger theorem. In the
presence of $h$, the latter reads:
\begin{equation}
Re \Sigma_{\sigma}(i0^{+}) - \mu = h\sigma - \mu_{0}(n_{\sigma})
\end{equation}
where $\mu_0$ denotes the chemical potential of the noninteracting system.
This condition expresses that the Fermi surface for each spin species
accomodates exactly $n_{\sigma}$ electrons.

In this regime of weak correlations, the Stoner approximation
provides a good description of our results for $m(h)$.
It is based on a Hartree-Fock decoupling of the interaction
with respect to $S_{z}$, leading to the RPA formula
for the uniform susceptibility
\mbox{$\chi/ \chi_{0} = [1-\frac{U}{2} \chi_{0}]^{-1} = S$}
($S$ is the so-called Stoner enhancement factor), and the
magnetization is given, at $T=0$, by:
\begin{equation}
m_{Stoner}= 2 \int_{0}^{h-m_{Stoner} U/2} d\epsilon \  D(\epsilon)
\end{equation}
This is compared to our results in Fig. \ref{mvsh(u=1)}
where the numerically obtained value of the ratio
$\chi(h=0)/\chi_0(h=0)\equiv S \simeq 1.74$ has been used as the input
parameter of the fit.
Within this approximation the low-temperature dependence of the
susceptibility reads:
\begin{equation}
\chi_{Stoner}(T) / \chi_{0} = S [1+\frac{\pi^{2}}{6}\frac{D''}{D}
S T^{2}+\cdots] = S [1-\frac{\pi^{2}}{12}S T^{2} + \cdots]
\label{chisto}
\end{equation}
from which a low-field dependence of the quasi-particle residue
can be inferred using Maxwell relation (\ref{Maxwell}) :
\begin{equation}
Z_{Stoner}(h)=Z(h=0) [1+\frac{h^2}{4} S^2 Z(h=0) + \cdots]
\label{zsto}
\end{equation}
This agrees very well with our results as shown on the
inset of Fig. \ref{mvsh(u=1)}.

Since the physics at larger $U$ is not that of a ferromagnetic instability
($\chi$ remains finite), we have not attempted to compare our results with
partial resummations of spin-fluctuation diagrams beyond Hartree-Fock theory
(as in the paramagnon approach \cite{PMAG}).

\subsection{The Mott Insulator}
As in the above case, the magnetization smoothly saturates for large $U$
($>>U_{c2}$)  with
$\partial \chi / \partial h < 0$ \  (Fig. \ref{mvsh(u=5)}), but the
crossover between the Mott insulator and the band insulator is now governed
by a different scale: the antiferromagnetic exchange $J=t^{2}/2 U$.
Because the scale
involved in {\it uniform} magnetic properties is the magnetic
exchange between one spin and a shell of $z$ antiparallel nearest-neighbor,
this scale survives the $z\rightarrow\infty$ limit.
This scale is absent in the Gutzwiller approximation, which yields $m(h)=1$
for all $h$ at $T=0$.

The Mott insulator has
$Re\Sigma(\omega+i0^+)\simeq 1/\omega$ at
$h=0$, but $h$ cuts-off this divergence, and a finite slope
at zero frequency is recovered for $h\neq 0$:
$Re\Sigma(\omega+i0^+)=\Sigma(i0^+)+(1-1/\alpha)\omega+\cdots$
(see Fig. \ref{sigmavsh(u=5)}).
This can be simply understood from the low-frequency limit of the
self-consistency relation (\ref{sc}) :
\begin{equation}
Re\Sigma_{\sigma}(\omega+i0^+)=\omega+\mu+h\sigma -
\frac{t^{2}}{2}ReG_{\sigma}(\omega+i0^+)-ReG^{-1}_{\sigma}(\omega+i0^+)
\end{equation}
with:
\begin{equation}
ReG_{\sigma}(\omega+i0^+)=-\int\!\!\!\!\! \backslash \ d\epsilon\
\frac{\rho_{\sigma}(\epsilon)}{\epsilon} - \omega
\int d\epsilon\  \frac{\rho'_{\sigma}(\epsilon)}{\epsilon} +\cdots
\equiv \rho_{1} + \omega \rho_{2} +\cdots
\end{equation}
Since, in presence of a finite field, the d.o.s. $\rho_{\sigma}(\omega)$
is not symmetric, the integral $\rho_{1}$ does not vanish, leading
to the linear slope of $Re\Sigma(\omega+i0^+)$, with:
\mbox{$1/\alpha=\rho_{2}/\rho^{2}_{1}-\rho_1 t^2/2$}.
$\alpha$
is a convenient measure of the crossover between the Mott insulator at
$h=0$ ($\alpha=0$) and the band insulator at large $h$ ($\alpha=1$),
and is depicted in the inset of Fig. \ref{sigmavsh(u=5)}.
The insulating (non-Fermi liquid) character of the solution all the
way from $h=0$ to saturation is evidenced by checking that
the Luttinger theorem is violated.

As expected, the magnetization curves are very poorly described by
a Stoner fit in this regime (adjusting the value of $S$,
saturation is found at a much smaller field $\simeq J$ in the actual
results than in the Stoner curve (Fig. \ref{mvsh(u=5)})). A quantitative
understanding of the results can nevertheless be reached by studying
the large $U$ limit of the fully connected model (\ref{rm}), along the
lines of ref.\cite{RKZ}.
For large $U$ and at half-filling, (\ref{rm}) reduces to an Hamiltonian
for spin degrees of freedom which reads:
\begin{equation}
H = \sum_{i j} J_{i j} \roarrow{S_{i}} \cdot  \roarrow{S_{j}}
\end{equation}
where $J_{i j}$ are independent random variable with
\mbox{$\overline{J_{i j}} = J/N$} and
\mbox{$\overline{J^{2}_{i j}}-\overline{J_{i j}}^{2} \propto J^{2}/N^{2}$}.
Since the variance is of order $1/N^2$ ({\it not} $1/N$),
the randomness becomes irrelevant in the thermodynamic limit. The
partition function of the pure model is easily obtained by the
steepest-descent method \cite{RKZ}, and the magnetization is simply given
by the Weiss mean-field equation:
\begin{equation}
m = \tanh{\beta [h - J m]}\ \ \ ,\ \ with \ \  J=\frac{t^{2}}{2U}
\label{mweiss}
\end{equation}
The uniform susceptibility also reads simply \cite{RKZ}:
$\chi (h=0) = 1/(T + J)$, while $\chi_{loc}=1/T$.

Figure \ref{mvsh(u=5)} shows the excellent agreement of expression
(\ref{mweiss}) with the numerical
results for $U=5$. This emphasizes the dominant role of the exchange
scale
$J$ in the Mott insulator, and the simplicity of its magnetic properties
in the $d=\infty$ limit.

Finally, the evolution of the
local density of states $\rho(\omega)=-\frac{1}{\pi} Im G(\omega+i0^+)$
obtained by exact diagonalization is consistent with the above scheme:
as the field is increased, the gap is practically not affected, but there is
an asymmetric redistribution of spectral weight (controlled by $J$)
between the high-energy peaks
corresponding to the upper Hubbard band.

\subsection{The Mott Transition and the strongly correlated Metal}
We studied intensively the $U=3$ case both: i) at $T=0.01$ (by complete
diagonalization with $n_{s}=4,5$) and
ii) at $T=0$ (by the Lancz\`{o}s method with $n_{s}=4,6,8$).
We also studied the value $U=3.5$ which lies inside the zero-field
coexistence region $[U_{c1},U_{c2}]$ at $T=0.01$.

The central result is that two coexisting solutions are found in the
interval
$h_{c1}(T) \leq h \leq h_{c2}(T)$ (see Fig. \ref{mvsh(u=3)}) :

-one obtained by decreasing the field adiabatically from the saturated,
high-field, regime

-one obtained by increasing $h$ adiabatically from the $h=0$ solution.

An unambiguous way to discriminate between these solutions
is to compare
the value of $Re\Sigma_{\sigma}(i0^{+})$ for each of them with the one
predicted by Luttinger theorem. This is done in the inset of
Fig. \ref{Zvsh(u=3)}
for the exact diagonalization results with $n_s=5$ at $T=0.01$.
Clearly, the solution obtained from the low-field one follows Luttinger
theorem in all the region where it exists. On the contrary, the
high-field one violates Luttinger theorem.
Thus, two different phases are present: a metallic (Fermi-liquid) one
obtained from the low-field regime, and an insulating (non-Fermi liquid)
one obtained from the high-field regime.

For the finite temperature results ($T=0.01$, $n_{s}=4,5$), the coexistence
region [$h_{c1}, h_{c2}$] at $U=3$ is found to be stabilized at the
values $h_{c1}\simeq 0.13, h_{c2}\simeq 0.17$ at $U=3$.
For $U=3.5$, $h_{c1}$ is lowered to $h_{c1}=0$ as expected from the
zero-field results.
The situation is not so clear for the $T=0$ (Lancz\`{o}s) results: as noted
in ref.\cite{MW}, a strong fluctuation of the field coexistence
interval with
the number of sites $n_{s}$ is found.

Our results allow to draw a rough estimate of the phase diagram in the
$(U,h)$ plane for a fixed low temperature (e.g $T=0.01$)
(see inset of Fig. \ref{mag(uvar)}).
A coexistence region
[$U_{c1}(h,T), U_{c2}(h,T)$] is found, which becomes narrower with
increasing field. The finite field, finite temperature Mott transition is
thus first-order: the precise location of the transition requires
a calculation of free-energies (Maxwell construction) which has not been
attempted in this work. The recent results of ref.\cite{RMK}
suggests that the first-order transition 'surface' $U_c(h,T)$
ends at a second-order critical point at $h=T=0$:
$U_c(h=0,T=0)=U_{c2}(h=0,T=0)$. Thus, the qualitative features of the
phase diagram are in rather good agreement with the predictions of the
Gutzwiller approximation: the Mott transition is indeed lowered and driven
first-order when a magnetic field is applied.

Finally, we characterize the physical properties of the strongly correlated
metal in a magnetic field (see Fig. \ref{mvsh(u=3)}).

The magnetization curve $m(h)$ starts linearly at $U=3$, and it shows
a metamagnetic curvature (${{\partial \chi}/{\partial h}}>0$) close to
the transition. This result has been obtained at $\beta =100$ and confirmed
as a function of the number of sites of the exact diagonalization procedure.

In comparison,
the Stoner prediction (with $S$ fitted from the $h=0$ slope) has an
appreciably larger downward curvature (faster saturation of $m(h)$), while
the Gutzwiller approximation yields too large an upward
curvature (thus predicting a stronger metamagnetism at low field than
what we find)
, and a first-order transition at
a lower magnetic field than what we find.
Similarly, the quasiparticle-residue $Z(h)$ at $U=3$ is practically
constant in the metal with a small decreasing tendency (increase in $m^*$)
near the coexistence
region (see Fig. \ref{Zvsh(u=3)}). Again, the
Gutzwiller approximation would predict too large an increase of the
effective mass upon increasing the field at low field. However, most of
the {\it qualitative} predictions of the Gutzwiller approximation are in
good agreement with our results.

Here also, we have found difficult to reliably estimate the low-temperature
dependence of
the zero-field susceptibility (i.e the sign of the $T^2$ correction).
However, the trend of the complete diagonalization results with
$n_s=4,5,6$ and the Lancz\`{o}s converged values
are compatible with a very flat $\chi(T)$, consistent with the
flatness of $Z(h)$ and Maxwell identity (\ref{Maxwell}).

Finally, the local density of states obtained in the coexistence region
is consistent with the characteristics of the two phases:
the metallic-like solution still have spectral weight
at the Fermi level, while the other one displays a gap.

\section{Concluding Remarks and Experimental Relevance}

We have studied in this paper the effect of a uniform magnetic field on
the paramagnetic solutions of the half-filled Hubbard model
in the limit of infinite dimensions. Depending on the interaction strength,
three regimes can be identified ({\it cf.} Fig.\ref{mag(uvar)}):

- In the weakly correlated metal at small $U$, the magnetic field
reduces the effect of the interaction because of the Pauli principle. A
smooth crossover is found between the unpolarized metal and the fully
polarized band insulator, with a mass enhancement $m^*/m$ decreasing
smoothly to unity. The field dependence of the magnetization at
zero-temperature is reasonably described by the Stoner formula.

- The uniform magnetic response of the Mott insulator at large $U$
is controlled by the antiferromagnetic exchange $J\simeq t^2/U$, while the
local susceptibility follows a Curie law. The field dependence of the
magnetization is described very well by a simple Curie-Weiss mean-field
expression.

- The most interesting magnetic properties are found at intermediate values
of $U$, close to the zero-field Mott transition. The applied magnetic
field induces a first-order metamagnetic transition between the strongly
correlated metal at low field and the high-field Mott insulator,
forcing
a jump in the magnetization curve. Similarly, the temperature drives
first-order the zero-field metal/insulator transition.
This is in qualitative agreement with the predictions of the Gutzwiller
approximation. Quantitatively however, this approximation does not describe
our results very well because it neglects the magnetic exchange. For
example, the critical field is predicted to be too low, and the upward
curvature of the $m(h)$ curve at low field much too large.

Finally, we discuss the experimental relevance of the present study
for the magnetic properties of transition metal oxides and liquid
$^3He$.

Ref.\cite{MR} reports measurements of the temperature dependence of
the susceptibility for $(V_{1-x}Cr_x)_2O_3$ for various $Cr$ concentrations
$x$. Sample $2$ of this paper ($x=0.008$) displays two phase transitions
as the temperature is raised:  from an antiferromagnetic insulator
to a paramagnetic metal (at $T_N\simeq 175K$), then form the metal to
a paramagnetic Mott insulator (at $T_{MI}\simeq 250K$). The measured
$\chi(T)$ is reproduced in part in the inset of Fig.\ref{chivsT(u=3.4)}.
Both transitions are clearly visible on this curve, even though the
second one is much broadened for reasons explained in
\cite{MR}. The Mott transition is signalled by a susceptibility increase.
Interestingly, $\chi(T)$ decreases with temperature both in the metal and
in the insulator above the transition region. Our treatment of the Hubbard
model can account qualitatively for this behavior, as
shown on Fig.\ref{chivsT(u=3.4)} displaying the uniform susceptibility
$vs.$ temperature found for $U=3.4$. The metallic solution must be followed
at low temperature, until the first order transition to the Mott insulator
is reached. Note that the temperature scale for the disappearance of the
metal is considerably lowered as compared to a typical electronic scale
of $t\simeq 10^4 K$, and indeed falls in the range of $10^2 K$.
Quantitatively however, the typical increase of $\chi$ through the Mott
transition is found to be larger than the experimental values even after
correcting for the observed broadening.

Regarding liquid $^3He$, the main conclusion of our work is that the
metamagnetic transition of the half-filled Hubbard model is not an
artefact of neglecting the exchange in the Gutzwiller approximation but
is indeed present in the more refined $d=\infty$ treatment. In spite of
the fact that we would predict an appreciably larger critical field for
this transition, our results for $m(h)$ are hardly compatible with
experiments \cite{WWP}. This means that a lattice description of liquid
$^3He$ by a Hubbard model rigidly maintained at half-filling is
not satisfactory. Introducing vacancies \cite{VWA} while keeping $U$ close
to the half-filled Mott transition might avoid metamagnetism,
but will still be
faced with a much too small susceptibility enhancement, of order $1/J$
({\it not} $1/T_K$) as soon as the exchange is correctly taken into account.
A lattice description, if at all possible, must find a way of suppressing
the exchange to restore the properties of a liquid ({\it e.g} by working at
very large $U$ as a function of vacancy concentration).

\acknowledgments

We acknowledge fruitful discussions with E. Wolf
on the polarized $^3He$ experiments,
and with M. T. Beal-Monod on paramagnon theory. We are grateful to
M. Caffarel for a previous collaboration on the exact diagonalization
algorithm and generous exchange of programs.
This work was supported by DRET contract $n\^{o} 921479$, and
L. Laloux is supported by a `Minist\`{e}re de la Recherche et de
l'Espace' contract.

\begin{figure}
\caption[fig1]{\\
1.a $\chi_{loc}$ {\it vs} $U$
at zero-field, for $\beta t=100$ ($n_{s}=5$).
A coexistence region between
$U_{c1}=3.3$ and $U_{c2}=3.8$ is clearly apparent.
\\
Inset:
\\
1.b -Z {\it vs} U: numerical solution (dots and curve) \cite{MW} and
Gutzwiller approximation (full line)
\\
1.c -Uniform susceptibility $\chi$ {\it vs} U for $\beta =100$ ($n_s=4$).
Note the linear dependence for large $U$, with slope
$\sim 1/J=2U/t^{2}$.}
\label{chilocvsu}
\end{figure}

\begin{figure}
\caption[fig2]{\\
2.a Magnetization curves $m(h)$ for, from below, $U=0,1,3,5,10$,
($\beta t=100$ and $n_{s}=4$).
Note the three regimes: metallic at small $U$ ($U=0,1$),
insulator at large $U$
($U=5,10$) and the transition region ($U=3$).
\\
2.b Inset: schematic phase diagram in the $(U,h)$ plane for T=0.01}
\label{mag(uvar)}
\end{figure}

\begin{figure}
\caption[fig3]{\\
$m$ {\it vs} h for $U=3$,
$\beta = 5,10,20,30,40,100$ (from below), and $n_{s}=4$.
The magnetic field induced phase transition occurs only at low temperature
($\beta \geq 50$).}
\label{mag(betavar)}
\end{figure}
\begin{figure}
\caption[fig4]{\\
4.a Magnetization $vs. h$  in the Gutzwiller approximation
for $U/U_{BR}=0,0.1,0.2,\cdots,0.9$ and also $U=2,3,4$ for
comparison with Fig. \ref{mag(uvar)}.
The Maxwell construction
has been done for $U > 0.44 U_{BR}$.
\\
4.b Inset: $(U,h)$ phase diagram at $T=0$ in the
Gutzwiller approximation: phase coexistence
region [$U(h_{c1}), U(h_{c2})$] (dotted lines), and first-order line
(full curve) ending at second-order points $(U_{BR},h=0)$ and
$(U_{c},h_{c})$.}
\label{mGutzvsu}
\end{figure}

\begin{figure}
\caption[fig5]{\\
5.a Magnetization of the weakly correlated metal $U=1$ ($\beta =100$ ,
$n_{s}=4$) (dashed curve). The Stoner prediction (full curve)
obtained by adjusting the Stoner factor
$S=\chi/\chi_{0}$ is in excellent agreement with the infinite-d result.
\\
5.b Inset:
Field-dependence of the quaasi-particle residue $Z(h)$ showing the
cross-over from the unpolarized to the fully polarized metal
(band insulator).
Low-field comparison to eq. (\ref{zsto})
(dashed curve).}
\label{mvsh(u=1)}
\end{figure}

\begin{figure}
\caption[fig6]{\\
Magnetization of the Mott insulator for $U=5$ ($\beta =100$ , $n_{s}=4$)
(small dashes+squares).
A Stoner fit ($S=\chi/\chi_{0}$) (lower curve) is very poor with a too
slow saturation of the magnetization. Conversely, a Curie-Weiss law
with $J=t^{2}/2U$ and $U=5$ (big dashes+circles), is in excellent
agreement with the infinite-d result.}
\label{mvsh(u=5)}
\end{figure}

\begin{figure}
\caption[fig7]{\\
7.a $Im\Sigma(i\omega_{n})$ {\it vs} $\omega$ at
($U=5$, $\beta =100$ and $n_{s}=4$) for various values of $h$ ($0.<h<.2$)
At $h\neq 0$, the divergence at $\omega=0$ is cut off,
leading to a finite slope
($1 - 1/\alpha(h)$) at small frequencies.
\\
7.b Inset: cross-over from Mott insulator to band insulator
(fully polarized) characterized by $\alpha(h)$.}
\label{sigmavsh(u=5)}
\end{figure}

\begin{figure}
\caption[fig8]{\\
Magnetization {\it vs} field at $U=3 $ ($\beta =100$ and $n_{s}=5,6$).
Middle curve: $d=\infty$ solution, exhibiting
$\partial m/ \partial h > 0$, and coexistence region
(2 solutions exist in the field interval
$0.13<h<0.17$). Stoner ($S=\chi/\chi_{0}$) (lower) and Gutzwiller
($U=3$) (upper) curves for comparison.}
\label{mvsh(u=3)}
\end{figure}

\begin{figure}
\caption[fig9]{\\
9.a Comparison of numerical solution with
Luttinger theorem prediction:
$Re\Sigma_{\uparrow}(i0^{+})-\mu$
$=h - \mu_{0}(n_{\uparrow})$.

The figure shows the two curves:
$m(h-Re\Sigma_{\uparrow}(i0^{+})-\mu)$
(dashed line) and $m(\mu_{0}(n_{\uparrow}))$
(full line), wich coincide
when Luttinger Theorem is satisfied.
The two curves differ for the insulating solution.
\\
9.b Inset: Evolution of $(1-\partial\Sigma/\partial\omega)^{-1}$
$vs. h$ ($Z(h)$ in the metal (full line), $\alpha(h)$ in the
insulator (dashed line)).
Note the small decreasing tendency of $Z(h)$ on
the metallic side close to the transition}
\label{Zvsh(u=3)}
\end{figure}

\begin{figure}
\caption[fig10]{\\
10.a Uniform susceptibility $\chi$ {\it vs} temperature for $U=3.4$
(in the $T=0.01$ coexistence region) calculated with $n_{s}=4$.
Lower curve (dotted): metallic solution. Upper curve (full):
insulating solution. A first-order metal/insulator transition
between the two branches takes place upon heating.
\\
10.b Inset: Results of ref.\cite{MR} for $(V_{1-x}Cr_{x})_{2}O_{3}$
(sample 2, $x=.008$). The first sharp rise corresponds to the Neel
temperature, followed by the (rounded) metal/Mott insulator transition}
\label{chivsT(u=3.4)}
\end{figure}


\begin{references}
\bibitem{foot2} The constant $U/2$ term is prescribed by the
Luttinger theorem $\Sigma(i0^+)=\mu=U/2$ (so that the
zero-frequency spectral density sticks to the non-interacting
value
$\rho(\omega=0)=\sqrt{2}/\pi t$ for metallic solutions at $h=0$).
\end{references}
\end{document}